\documentclass{article}

\usepackage{PRIMEarxiv}

\usepackage[utf8]{inputenc} 
\usepackage[T1]{fontenc}    
\usepackage{hyperref}       
\usepackage{url}            
\usepackage{booktabs}       
\usepackage{amsfonts}       
\usepackage{nicefrac}       
\usepackage{microtype}      
\usepackage{lipsum}
\usepackage{fancyhdr}       
\usepackage{graphicx}       
\graphicspath{{media/}}     
\usepackage{amssymb,amsfonts}
\usepackage{xcolor}  

\title{R\lowercase{o}TIR: Rotation-Equivariant Network and Transformers for Fish Scale Image Registration
}

\author{
  Ruixiong Wang\thanks{\textbf{Corresponding author}: ORCID: 0000-0003-1824-185X}
  , Alin Achim, Renata Raele-Rolfe, Qiao Tong, Dylan Bergen, Chrissy Hammond, Stephen Cross
  \\
  University of Bristol \\
  Bristol, United Kingdom\\
  \texttt{\{ruixiong.wang, alin.achim,..., stephen.cross\}@bristol.ak,uk} \\
}

\begin{document}

\noindent
\textcolor{red}{\textbf{Note:}} \textcolor{red}{This is the initial version of the work that has been published in the 28th UK Conference on Medical Image Understanding and Analysis (MIUA 2024) with the revised title ``RoTIR: Rotation-Equivariant Network and Transformers for \textbf{Zebrafish} Scale Image Registration''. Please refer to the published version, DOI: \href{https://doi.org/10.1007/978-3-031-66955-2_20}{10.1007/978-3-031-66955-2\_20}. The presentation is available on YouTube (\href{https://www.youtube.com/watch?v=wc4t63IJFiE&list=PLFm8siRFfuk1N__sqp0EkgIzrecF-YW7i}{RoTIR presentation video}).}

\vspace{10pt} 

\maketitle

\begin{abstract}
Image registration is an essential process for aligning features of interest from multiple images. With the recent development of deep learning techniques, image registration approaches have advanced to a new level. In this work, we present 'Rotation-Equivariant network and Transformers for Image Registration' (RoTIR), a deep-learning-based method for the alignment of fish scale images captured by light microscopy. This approach overcomes the challenge of arbitrary rotation and translation detection, as well as the absence of ground truth data. We employ feature-matching approaches based on Transformers and general E(2)-equivariant steerable CNNs for model creation. Besides, an artificial training dataset is employed for semi-supervised learning. Results show RoTIR successfully achieves the goal of fish scale image registration. 
\end{abstract}

\keywords{Image registration \and feature matching \and steerable CNNs \and rotation equivariance \and transformers}

\section{Introduction}
\label{sec:intro}

The primary objective of image registration is to calculate the transformation functions that align one image to another. For this, the affine transformation is commonly utilised, which can account for operations such as translation, rotation and scaling \cite{wang2014robust}. To determine the transformation, extraction of common features in the two images is a fundamental task \cite{kuppala2020overview}. Deep learning, specifically using convolutional neural networks (CNNs), is an efficient approach to implementing hidden feature extraction and has already been introduced in multiple computer vision tasks. 

In our study, our objective was to develop an algorithm for registering images of zebrafish fish scales cultivated in Petri dishes. Scales are mini bone organs with bone-forming and -resorbing osteoclast which have been used in the study of skeletal regeneration and identification of bone anabolic compounds \cite{bergen2019zebrafish}. Correct registration will enable better data collection. Moreover, registration can also be used in the study of monitoring bone healing after small bone injuries \cite{dietrich2021skeletal}. However, due to motion from the routine addition of media, the positions of the fish scales changed over time presenting a challenge for fish scale observation.

It is normally agreed that there is no approach that is generally applicable to all image registration conditions \cite{fischer2008ill}. As such, the fish scale images used in this work, have their own peculiarities. The transformations are mainly confined to translation and rotation, while the effects of scaling and warping are limited. Furthermore, the images used contain two features which behave independently: the edge of the well in which the fish scale is placed is typically static, while the fish scale itself can undergo large positional shifts within this space. 
For training, the only resources available are the raw microscope images with no ground truth that can be used for supervision or performance evaluation. Thus, building training datasets and establishing an appropriate loss function are crucial \cite{chen2021deep}. 

In this work, we create the RoTIR model that employs the theory of general E(2)-equivariant steerable CNNs for feature extraction \cite{weiler2019e2cnn} and Transformer-based modules for feature-point matching \cite{sun2021loftr}. Additionally, we generate an artificial training dataset to support semi-supervised learning. The paper is organized as follows: Section \ref{sec:background} provides an overview of the background knowledge that forms the foundation of our work. In Section \ref{sec:approach}, we delve into the details of the algorithms employed, covering aspects such as network architecture, the training dataset, and the loss function. The registration results produced by RoTIR are discussed in Section \ref{sec:result}. Finally, Section \ref{sec:conclusion} encapsulates and summarizes the outcomes of the study.

\section{Background}
\label{sec:background}

\subsection{Siamese Networks for Template Matching}

Template matching is a similarity learning process that identifies an arbitrary exemplar image pattern within a larger search image. Siamese network-based models employ feature extraction networks with identical architectures and parameters for both templates and targets and identify all potential matching locations. Subsequently, similarity learning techniques, such as cross-correlation, are applied to determine the best matching position. Liu \textit{et al.} introduced the rotation-invariant methods to the Siamese network for tracking and developed the RE-SiamNet \cite{liu2020rotation} which received better rotation detection.

\subsection{Feature-point Matching Model Based on Transformers}

Template-matching methods emphasize global feature alignments, whereas approaches of feature point matching concentrate on the alignment of local features. The procedure of feature-matching models can be summarized into four steps: detection, description, matching and filtering \cite{bokman2022case}. Recently deep-learning-based methods have achieved significant performance. Sun \textit{et al.} introduced their Local Feature TRansformers (LoFTR) model \cite{sun2021loftr} which was a detector-free approach that directly generated dense feature fields for matching. After that, Bokman \textit{et al.} substituted the backbone of LoFTR with group equivalent neural networks and received better applicability \cite{bokman2022case}.

\subsection{Rotation-equivariant Neural Networks as Backbone}

Both the work of Liu \textit{et al.} \cite{liu2020rotation} and Bokman \textit{et al.} \cite{bokman2022case} introduced the rotation equivariant neural networks into the existing models. A brief interpretation of the transformation equivariant neural networks could be expressed as $F[T^{in}(I)] = T^{out}[F(I)]$, where $I$ is the input, $F$ is the function, and $T$ is the inside or outside transformation function \cite{bokman2022case}. The framework of general E(2)-equivariant steerable CNNs holds the equation above for translation and rotation.

Steerable CNNs are a type of group equivariant neural networks that impose constraints on convolutional kernels using group representation theory, which defines the transformation laws within feature spaces \cite{cohen2019general}. The framework of CNNs exhibits equivariance under the Euclidean Group $E(2)$ and consists of two components: the translation group $(\mathbb{R}^2, +)$ and the orthogonal group $O(2)$, encompassing rotations and reflections \cite{weiler2019e2cnn} which are combined through the semi-product operation, yielding $E(2)\cong(\mathbb{R}^2,+) \rtimes O(2)$. 

Feature spaces of steerable CNNs are defined as spaces of feature fields. They are characterized by the group representation that determines the transformation functions for the inputs. Steerable feature fields are presented by $f: \mathbb{R}^2 \rightarrow \mathbb{R}^c$, where $c$ is the dimension of the feature vectors corresponding to the pixels on the image plane $\mathbb{R}^2$ \cite{weiler2019e2cnn}. Therefore, the feature vectors within the steerable feature fields possess orientation-specific characteristics.  

\section{Proposed Approach}
\label{sec:approach}

\subsection{Network Architecture}

\subsubsection{Backbone Network}

\begin{figure}[tb]
\centerline{\includegraphics[width=0.65\linewidth]{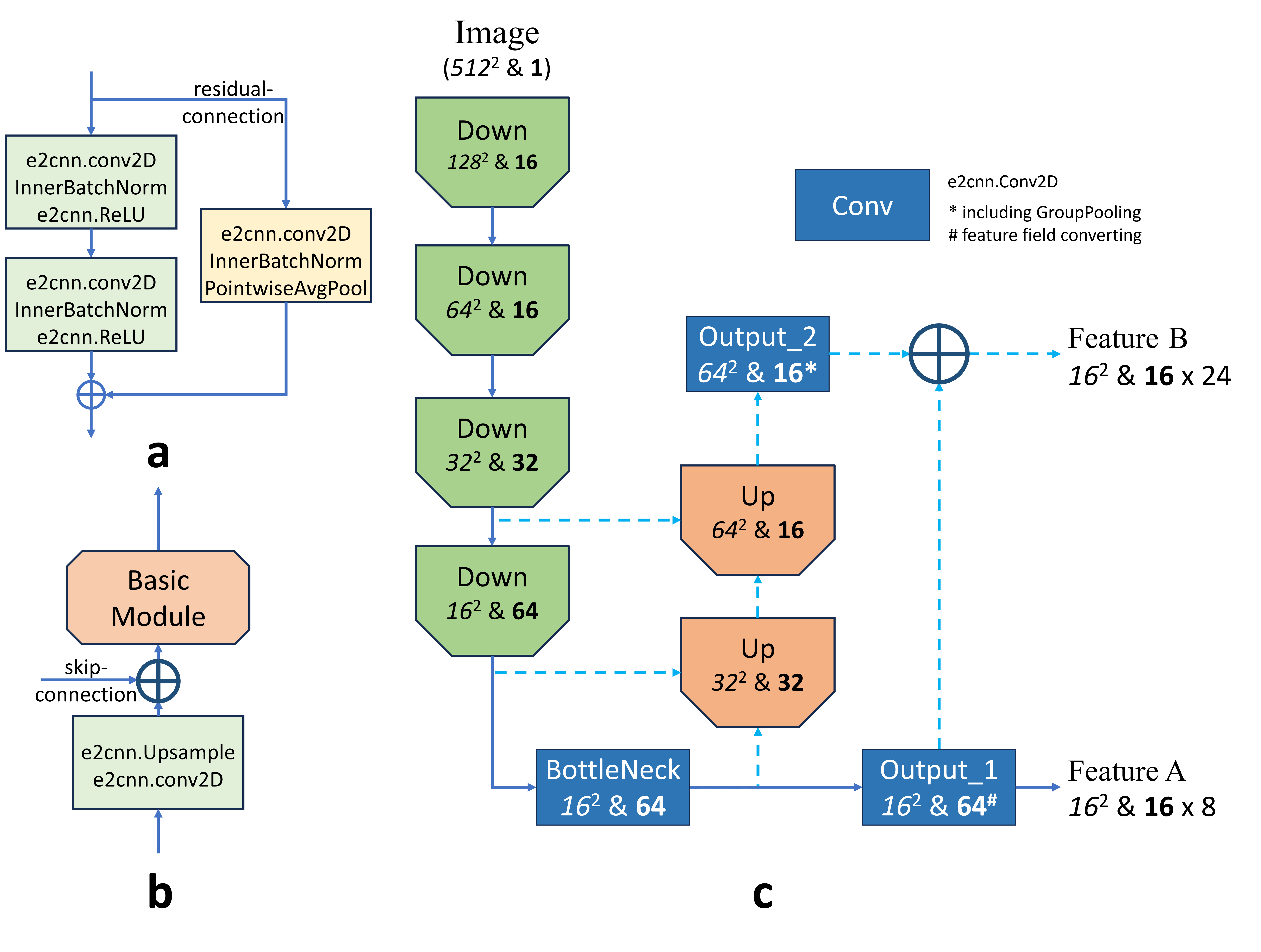}}
\caption{Architecture of Backbone Network. (\textbf{a}) Basic module, down-sampling or size-unchange process, (\textbf{b}) up-sampling module, contain size-unchange basic module, (\textbf{c}) backbone network.}
\label{fig:backbone}
\end{figure}

The backbone network aims for feature extraction. It derives from the classical object detection model YOLO \cite{redmon2016you} and transforms the input image into an $S \times S$ grid-shaped feature field with each pixel representing a patch at the corresponding position of the input image. The advantage of this design is that it bypasses the need for specific feature point extraction and selection. The architecture of the backbone network is shown in Figure \ref{fig:backbone}. The fundamental process of the backbone network is down-sampling, which outputs a feature field with sizes of $16 \times 16$. The backbone network is constructed using E(2)-equivariant CNNs, the inputs, hidden features and outputs are transformed into scalar, regular, vector feature fields. The down-sampling output has 4 vector fields, as the vector feature fields involve two dimensions and 8 output channels. 

Inspired by the LoFTR model \cite{sun2021loftr}, we expand the backbone network with an up-sampling process after the down-sampling process. The output feature fields of the up-sampling process, under regular representation, produce outputs 4 times the size of the down-sampling outputs. Finally, outputs are rearranged to 16 channels. The outputs from down- and up-sampling are concatenated together with 24 channels as the extracted features for matching.

\subsubsection{Matching Module}

\begin{figure}[htbp]
\centerline{\includegraphics[width=0.6\linewidth]{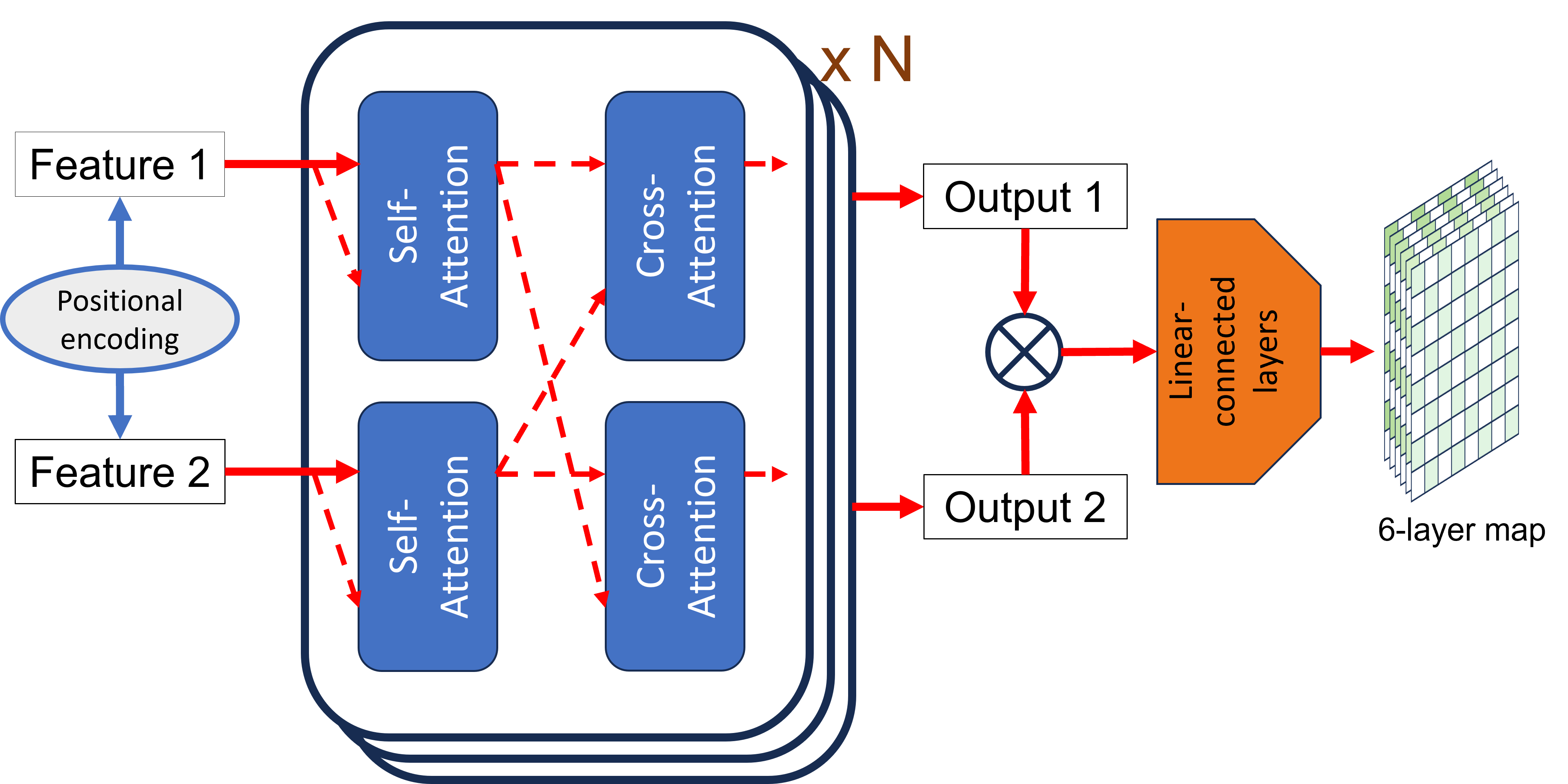}}
\caption{Architecture of matching module.}
\label{fig:matchingmodule}
\end{figure}

The matching module relies on the local feature transformer module in the LoFTR model. A pair of features extracted from the backbone network and added with positional encoding and then are flattened and fed into the matching module for similarity computation and matching. The module is comprised of a series of alternately permuted self- and cross-attention layers. For computational efficiency, the attention layers in the RoTIR model are linear attention. One difference to the LoFTR module is the number of output channels. Our RoTIR model intends to provide independent parameters for each transformation term. Thus, we add a linear projection head after the matching module converting the outputs to 6 channels, then translating them into a matching map and parameters for rotation angles, scaling factor and coordinate refinement. The architectures of the matching module and output head are shown in Figure \ref{fig:matchingmodule}.

\subsection{Training Dataset Synthesis}

We used separate parameters for the affine transformation matrices. Therefore, a simulated training dataset using raw images with labels for all parameters was prepared \cite{xia2019convolutional}. The process of building the synthetic dataset mainly contained three aspects: raw image cropping, image pair synthesis and ground-truth production. By image cropping, the foreground fish scales were isolated from the backgrounds. Next, random parameters for translation, rotation and scaling were applied to form the moving and fixed images. These same selected parameters were then used to produce the transformation matrices. As the backbone networks split the images into $16 \times 16$ patches, the matching maps in the generated ground truth have shapes of $256 \times 256$.

\subsection{Loss Function}

The total loss of the system is computed as weighted sums of independent terms. The most essential part is the calculation of the loss for confidence maps. Sinkhorn iteration is used in rearranging the confidence map, with the processed map containing one more column and row as bin storage for instances of no matches \cite{cuturi2013sinkhorn}. The loss for confidence maps is calculated using negative log-likelihood loss. The rotation angles are represented by their sine and cosine values. Coordinate refinement refers to the adjustment of the locations of matched points of interest on fixed images. If a scaling factor is involved, an exponent value with a base of 1.5 is used for loss calculation. The losses for angles, translation refinement and scaling factors are calculated using the L2 loss respectively.

\section{Results}
\label{sec:result}

For assessment, 45 pairs of phase contrast images with corresponding fish scale masks were used. Images were acquired using an Incucyte Zoom high-content live cell imaging system (Sartorius AG) and 4x lens (0.2 NA). The masks are used for evaluation metrics (e.g. DICE index) and background noise removal.

\subsection{Evaluation Metrics}

The DICE similarity coefficient (DICE) index and complex wavelet structural similarity (CW-SSIM) index are used for registration assessment. The DICE index serves as an intuitive similarity assessment relying on the Intersection over Union (IoU) of the targets. SSIM emulates the prediction of image quality by human visual perception. It offers significant performance in image similarity qualification with low computational complexity but has a high sensitivity to rotation and scale distortions. CW-SSIM is an extension of the SSIM method to the complex wavelet domain. It separates the measurement of magnitude and phase distortions and emphasises consistent relative phase distortions \cite{sampat2009cwssim}. The CW-SSIM index was insensitive to small distortions and provided gradual variance values for continuous distortions.

\subsection{Training Details}

The code was implemented using Python and PyTorch. The learning rate in the training process was set to $0.001$, with the Adam optimizer. Three suffix letters are used to indicate the configuration of RoTIR models. The first two are for usages of the up-sampling process in the backbone network and scale detection, (\textbf{T} for true, \textbf{F} for false), while the last refers to the application of rectangular masks identifying the coarse location of fish scales, (\textbf{T} for applied, \textbf{F} for not).

\subsection{Rotation Detection Robustness}

\begin{table}[htbp]
\caption{Rotation Angle Detection}
\begin{center}
\begin{tabular}{c c c c c c c}
\hline
Model &  Max$^{\mathrm{a}}$ & \multicolumn{5}{c}{Rotation angles at Max Std.$^{\mathrm{b}}$}\\
\cline{3-7}
RoTIR  & Std.$^{\mathrm{ }}$ & $0^{\circ}$ & $90^{\circ}$ & $180^{\circ}$ & $270^{\circ}$ & Ext.$^{\mathrm{c}}$ \\
\hline
TTF   & 8.14  & $123^{\circ}$ & $128^{\circ}$ & $112^{\circ}$ & $131^{\circ}$ & $18.5^{\circ}$ \\ 
TFF   & 5.97  & $254^{\circ}$ & $248^{\circ}$ & $262^{\circ}$ & $251^{\circ}$ & $13.7^{\circ}$ \\ 
FTF   & 3.68  & $272^{\circ}$ & $269^{\circ}$ & $274^{\circ}$ & $266^{\circ}$ & $8.23^{\circ}$ \\ 
FFF   & 5.66  & $268^{\circ}$ & $269^{\circ}$ & $269^{\circ}$ & $257^{\circ}$ & $12.0^{\circ}$ \\ 
\hline
\multicolumn{7}{l}{$^{\mathrm{a}}$Std. represents standard deviation.}\\
\multicolumn{7}{l}{$^{\mathrm{b}}$Rotations for moving images have already been subtracted }\\
\multicolumn{7}{l}{\hspace{0.75em}from detected angels.} \\
\multicolumn{7}{l}{$^{\mathrm{c}}$Ext. represents extreme value.}\\
\end{tabular}
\label{tab:rotationanaly}
\end{center}
\end{table}

We found that the variance of DICE and CW-SSIM indexes was unpredictable when the rotation angles exceeded $90^{\circ}$. Therefore, we began by testing the registration for rotation detection to ensure that the output rotation angles fell within the valid reasonable range for evaluation. The process was achieved by freezing the fixed images and rotating the moving images by $0^{\circ}$, $90^{\circ}$, $180^{\circ}$ and $270^{\circ}$. Ideally, values of detected angles subtracting the rotations of moving images should be the same. Table \ref{tab:rotationanaly} shows the maximum standard deviations of the 45 groups under four testing models. From the table, we can see that detected angle errors fell within an acceptable range, demonstrating that the rotation detection capability of the RoTIR model was robust and our evaluation metrics were feasible in quantifying the testing result.

\subsection{Evaluation of the Registration Result}

\begin{figure}[htbp]
\centerline{\includegraphics[width=0.7\linewidth]{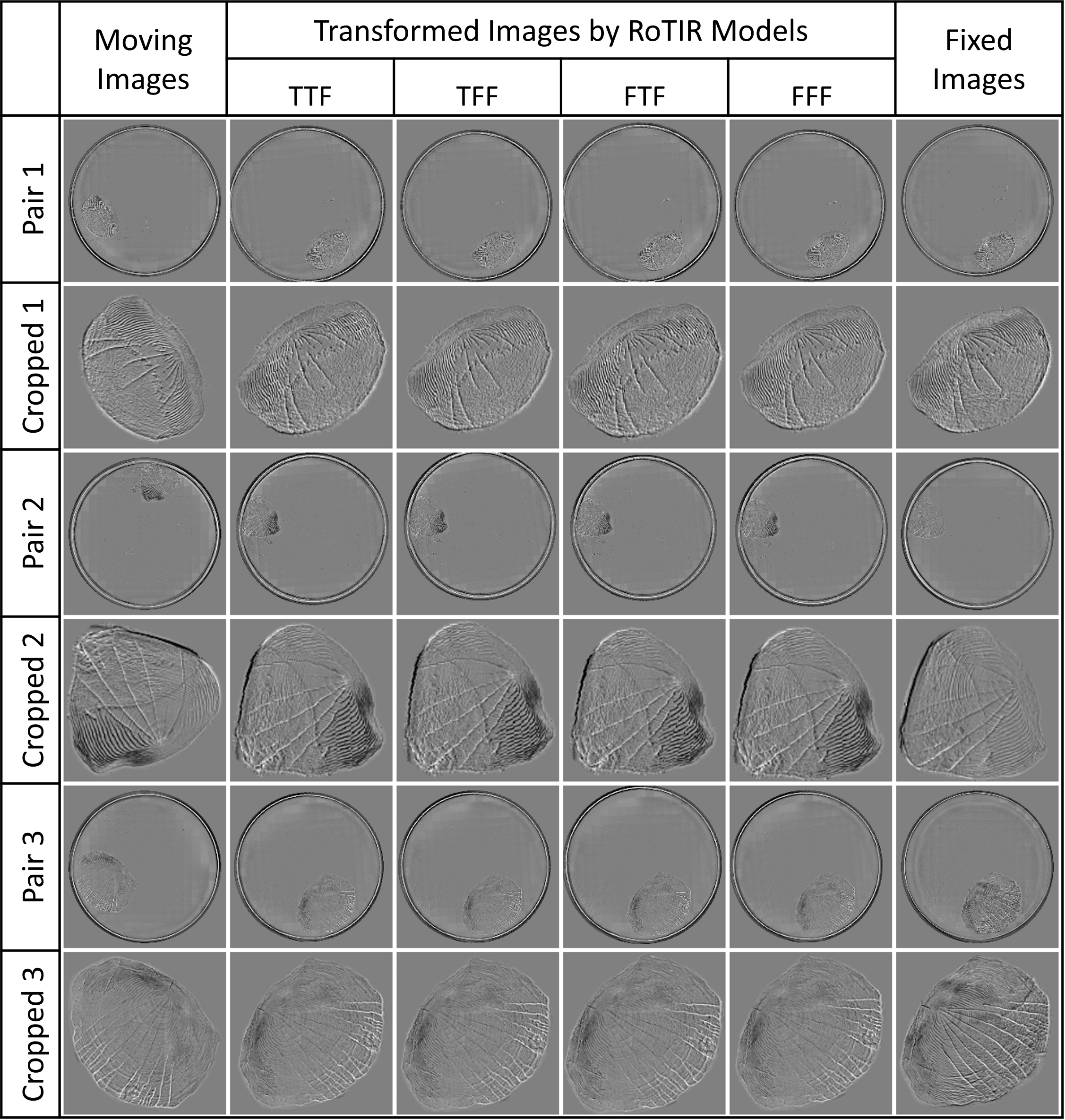}}
\caption{Registration results by RoTIR models.}
\label{fig:registrationresult}
\end{figure}

The results for RoTIR models computed with the evaluation metrics are shown in Table \ref{tab:registrationresult}. For each model, we computed the mean and standard deviation values of DICE and CW-SSIM indexes for all 45 pairs of images with 4 direction rotations. All versions of the RoTIR models tested achieved DICE indexes higher than 0.9, and most of the CW-SSIM indexes exceeded 0.4. Figure \ref{fig:registrationresult} shows four example registrations.

\begin{table}[htbp]
\caption{Registration result of RoTIR Model}
\begin{center}
\setlength{\tabcolsep}{1.5em}{
\begin{tabular}{l c c}
\hline
\multicolumn{1}{c}{Model}  &  DICE  &  CW-SSIM  \\
\hline
RoTIR\_TTF & 0.937 $\pm$ 0.025 & 0.420 $\pm$ 0.035 \\
RoTIR\_TTF* & 0.922 $\pm$0.036 & 0.407 $\pm$ 0.039 \\
RoTIR\_TTT & 0.942 $\pm$ 0.022 & 0.423 $\pm$ 0.035 \\
\hline
RoTIR\_TFF & 0.934 $\pm$ 0.035 & 0.419 $\pm$ 0.038 \\
RoTIR\_TFF* & 0.916 $\pm$ 0.044 & 0.404 $\pm$ 0.038 \\
RoTIR\_TFT & 0.935 $\pm$ 0.030 & 0.419 $\pm$ 0.036 \\
\hline
RoTIR\_FTF & 0.925 $\pm$ 0.042 & 0.411 $\pm$ 0.035 \\
RoTIR\_FTF* & 0.914 $\pm$ 0.045 & 0.400 $\pm$ 0.031 \\
RoTIR\_FTT & 0.936 $\pm$ 0.040 & 0.420 $\pm$ 0.038 \\
\hline
RoTIR\_FFF & 0.944 $\pm$ 0.027 & 0.422 $\pm$ 0.038 \\
RoTIR\_FFF* & 0.925 $\pm$ 0.040 & 0.404 $\pm$ 0.040 \\
RoTIR\_FFT & 0.945 $\pm$ 0.025 & 0.422 $\pm$ 0.039 \\
\hline
\multicolumn{3}{l}{* Results that do not utilize coordinate refinement.} \\
\end{tabular}}
\label{tab:registrationresult}
\end{center}
\end{table}
 

We evaluated the impact of coordinate refinement of the RoTIR models. The confidence maps indicated the pairs of matching points from moving and fixed images. However, confidence maps are only capable of indicating the centroids of image patches, not precise locations. Coordinate refinement aims to compute the compensations to the centroids on the fixed image. From Table \ref{tab:registrationresult} we can see that models with coordinate refinement all performed better in DICE and CW-SSIM in every situation. 


Spatial scale detection is another key phase in some registration tasks; however, the size differences in our dataset were minimal. The RoTIR\_FFN model received the best performance (Table \ref{tab:registrationresult}), yet this didn't involve scale detection. Likewise, scaling factors predicted by RoTIR\_TTN and RoTIR\_FTN were just $1.001 \pm 0.004$ and $1.002 \pm 0.008$.  As such, scale detection capability is not fully proven. Additionally, the use of rectangle masks, which coarsely identified the locations of fish scales, did not show a significant improvement, especially for models without scale detection. Considering the computational efficiency of mask generation, the application of rectangle masks was not essential for our task.

\subsection{Key Point Detection}

\begin{figure}[htbp]
\centerline{\includegraphics[width=0.85\linewidth]{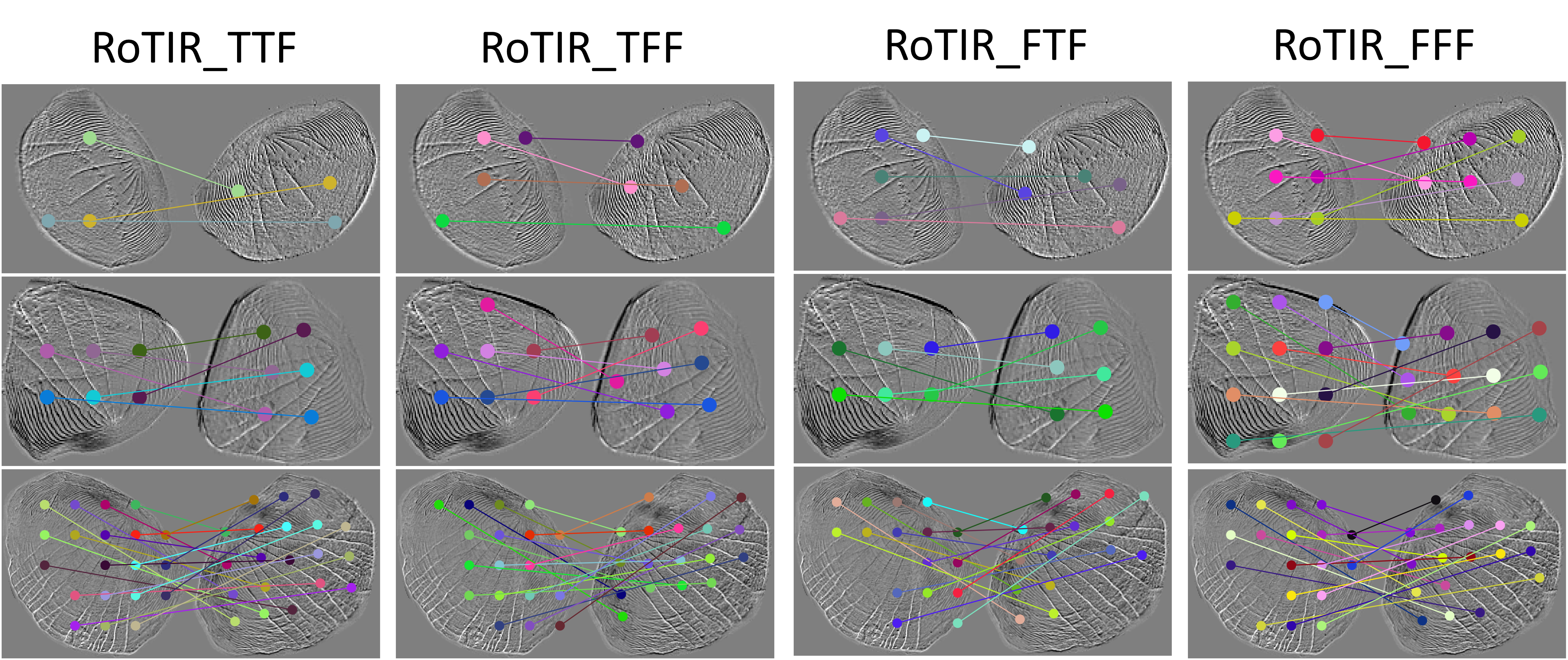}}
\caption{Key point matching results by RoTIR models.}
\label{fig:keypointmatching}
\end{figure}

Our model used the centroids of gridded patches as potential feature key points, therefore points of interest detected in moving images were neatly lined up. However, locations of corresponding matching points on fixed images were not restricted to centroids because of coordinate refinement. The matching results from Figure \ref{fig:registrationresult} are shown in Figure \ref{fig:keypointmatching}. We see, that our matching strategy was well-suited for our situation and yielded reasonable feature matching and image registration results.

\section{Conclusion}
\label{sec:conclusion}

The RoTIR model developed in this study effectively and efficiently met the demands for the image registration task for a series of fish scale images. We have overcome the challenge of a lack of ground truth and successfully implemented the detection of arbitrary rotation. The general E(2) equivariant steerable CNNs provided a decisive contribution to rotation angle detection. Meanwhile, the matching algorithm based on Transformers provided reliable similarity alignment outcomes. Our RoTIR model successfully eliminated the restriction of rotation angles and translation distance. However, due to insignificant scale variance, our RoTIR models did not have the chance to exhibit their capability in scale detection. The applicability of RoTIR to more general image registration tasks poses an interesting avenue for further study. 

\section{Compliance with ethical standards}
\label{sec:ethics}

All zebrafish experiments were approved by the local Animal Welfare and Ethical Review Board (Bristol AWERB), and were conducted under a Home Office Project Licence.

\section*{Acknowledgments}
\label{sec:acknowledgments}

This work was carried out using the BlueCrystal Phase 4 facility of the Advanced Computing Research Centre, University of Bristol. We thank the Wolfson Bioimaging Facility (Bristol) for providing access to the Incucyte Zoom live cell imaging system.

\bibliographystyle{unsrt}  
\bibliography{references}

\end{document}